# Evidence for ion migration in hybrid perovskite solar cells with minimal hysteresis

*Philip Calado[1§], Andrew M. Telford[1§], Daniel Bryant[2,3], Xiaoe Li[3], Jenny Nelson[1,3], Brian C. O'Regan[4**], Piers R. F. Barnes[1*]*

1    Department of Physics, Imperial College London, SW7 2AZ, UK
2    Department of Chemistry, Imperial College London, SW7 2AZ, UK
3    SPECIFIC, Swansea University, SA12 7AX, UK
4    Sunlight Scientific, 1190 Oxford Street, Berkeley CA, 94707, USA

§    These two authors have contributed equally to this work
*    piers.barnes@imperial.ac.uk
**   bor@borski.demon.co.uk

**Abstract**

Ionic migration has been proposed as a possible cause of photovoltaic current-voltage hysteresis in hybrid perovskite solar cells. A major objection to this hypothesis is that hysteresis can be reduced by changing the interfacial contact materials, which are unlikely to significantly influence the behaviour of mobile ionic charge within the perovskite phase. Here we use transient optoelectronic measurements, combined with device simulations, to show that the primary effects of ionic migration can in fact be observed both in devices with hysteresis, and with 'hysteresis free' type contact materials. The data indicate that electric-field screening, consistent with ionic migration, is similar in both high and low hysteresis $CH_3NH_3PbI_3$ cells. Simulation of the transients shows that hysteresis requires the combination of both mobile ionic charge and recombination near the contacts. Passivating contact recombination results in higher photogenerated charge concentrations at forward bias which screen the ionic charge, reducing hysteresis.

Lead-halide perovskite solar cells have recently emerged as a promising new solution-processable photovoltaic technology. Despite rapid developments in efficiency figures and fabrication processes, the performance of many of these cells remains strongly dependent on their prior optical and electronic conditioning.[1-3] This was first noted as an 'anomalous hysteresis' in the characteristic current-voltage (*J-V*) scan in $CH_3NH_3PbI_3$ devices (known herein as simply 'hysteresis').[2] Subsequently the same process has been measured as a slow change in device photocurrent, photoluminescence intensity, and open circuit voltage ($V_{oc}$) occurring on timescales up to hundreds of seconds,[3-9] and its magnitude tends to increase with aging/degradation.[3,10-13] Understanding this mechanism and its effect on photovoltaic performance is critical for directing future perovskite solar cell research, to either resolve the issue, or exploit it.[7,14,15]

The leading model to explain hysteresis is that time-varying quantities of charge accumulated at the $CH_3NH_3PbI_3$ interfaces reduces or entirely screens the internal electric field, resulting in loss of photocurrent.[1,6,7,16-20] Migration of ionic defects in the perovskite phase, ferroelectric polarisation, or trapping of electrons at the interfaces have all been suggested as mechanisms for the accumulation of this charge.[2] However, it is not clear that ferroelectric polarisation can persist in these materials.[21,22] The rate of any ferroelectric polarisation/depolarisation or the rate of trapping/detrapping of electrons at interfacial states would be likely to be fast[17,21,23] relative to the slow timescales related to hysteresis phenomena (1 – 100 seconds) to be solely responsible.[1-3,5,6] There is strong direct and indirect evidence that slow drift and diffusion of ionic defects at room temperature is the dominant mechanism underlying hysteresis in $CH_3NH_3PbI_3$ solar cells.[6,7,17,24-31] However, a significant objection to this hypothesis is that the degree of observed hysteresis is highly dependent on the interface properties and choice of contact materials; these appear to control the interfacial trap density.[2,19,32-40] For example when a ZnO cathode top layer is replaced with phenyl-C61-butyric acid methyl ester (PCBM) in a $CH_3NH_3PbI_3$ device with an otherwise identical architecture, hysteresis at room temperature is significantly reduced (see figures S1and 1c). To a first approximation, ionic defect concentration and mobility in the bulk of the perovskite phase are not expected to be strongly influenced by the contacts, although the possibility that PCBM blocks ionic migration at grain boundaries has been proposed.[30] These observations undermine the viability of the ionic diffusion model for explaining hysteresis because the effect appears to be controlled by the contact material properties rather than the accumulation of mobile ions in perovskite layer at the interfaces.

Recent simulations suggest that *J-V* hysteresis could only be reproduced accurately if both ionic migration and recombination via interfacial traps were present in devices.[16] Here we present transient optoelectronic measurements that probe the direction of the internal electric field in operating devices. The measurements directly indicate that ionic migration appears in devices both with and without hysteresis. Our simulations reproduce the transient device behaviour over all relevant timescales ($10^{-8} - 10^2$ s). The results show that hysteresis is only observed in cases where high rates of recombination exist in the perovskite/contact interfacial regions of devices. During the forward *J-V* scan a reverse electric field in the bulk perovskite layer drives electrons and holes away from their respective transport layers and high concentrations of minority carriers build up at these interfaces. Where these interfaces act as recombination regions, the charge collection efficiency of the device is adversely affected. When recombination at these interfaces is reduced, instead, high photogenerated carrier concentrations at these interfaces contribute to efficient collection of diffusive currents during the forward scan. Low hysteresis is thus primarily an artefact of low interfacial recombination and resultant high photogenerated carrier populations at forward bias, despite the presence of ionic migration. Our evidence experimentally confirms the prediction of van Reenen et al.[16] and indicates that the stability of photocurrents and photovoltages in a device can be controlled by changing the interfacial recombination properties, without necessarily requiring a change in ion concentration or mobility within the perovskite phase. These observations resolve a significant concern for our understanding of hysteresis in this material.

**Experimental Results**

We examined $CH_3NH_3PbI_3$ solar cells with two device architectures (see Methods for details): one showing very limited *J-V* hysteresis at room temperature (herein referred to as 'top cathode' cells – see figures 1a and 1c) and the other showing significant hysteresis in the photovoltaic performance at room temperature (referred to as 'bottom cathode' cells – see figures 1b and 1d).

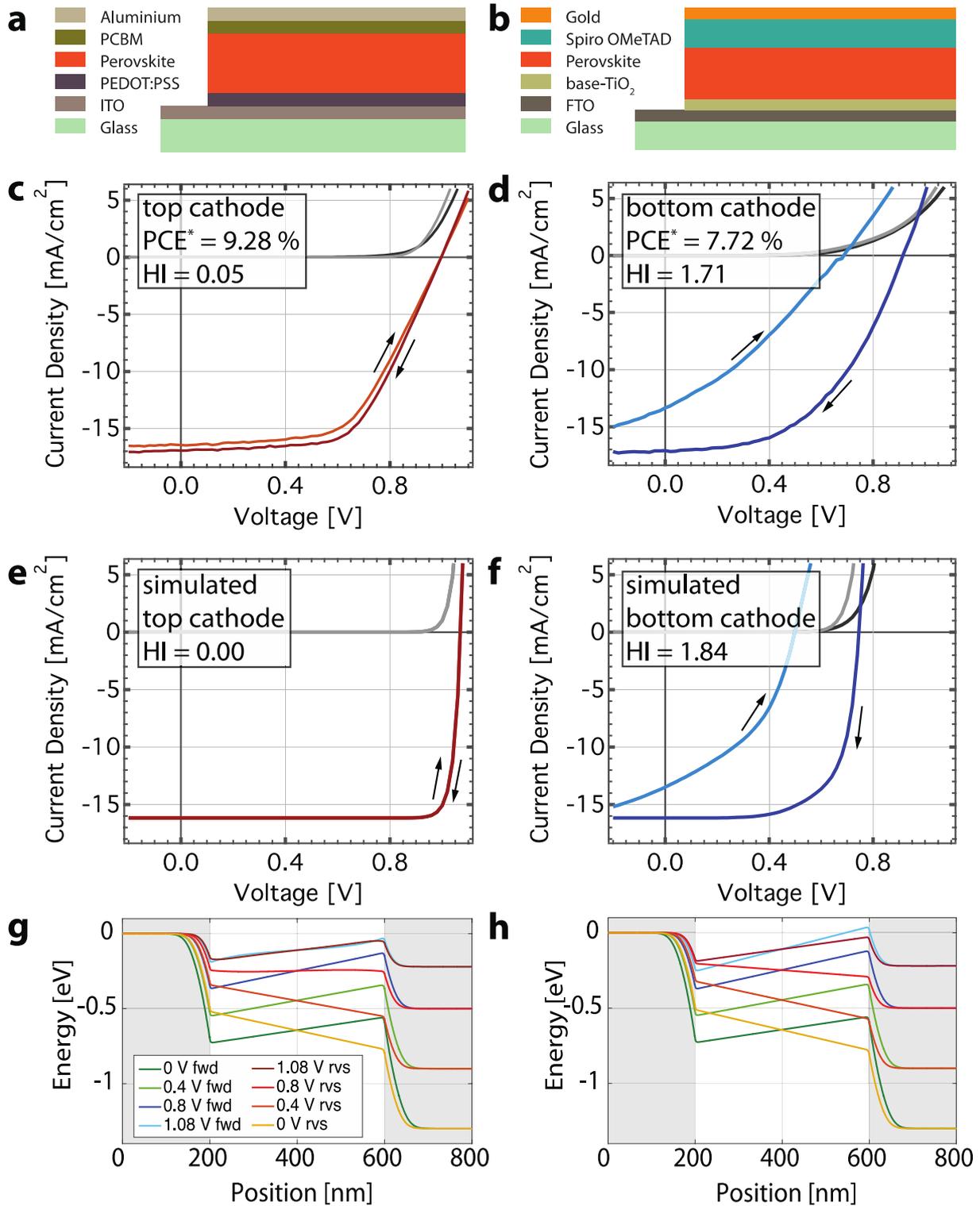

**Figure 1.** **Measured and simulated device current-voltage characteristics.** (**a**) Top cathode and (**b**) bottom cathode perovskite solar cell device architecture stacks. Measured current-voltage curves in the dark and under '1 sun' illumination scanned at approximately 40 mVs$^{-1}$ in the forward and reverse bias directions for the (**c**) top cathode and (**d**) bottom cathode architecture. *Power conversion efficiencies (PCE) calculated from the reverse scans (from forward to reverse bias), the hysteresis index

(HI) is defined in the methods. The corresponding simulated current-voltage scans in each scan direction are shown for a p-i-n device structure with mobile ions without (**e**) and with (**f**) recombination in the p- and n-type contacts. The scan protocol was similar to the experimental measurement at 40 mVs$^{-1}$ (see Methods for details); the effects of series resistance were not included which accounts for the discrepancy in fill factor between experiment and simulation. Simulated electrostatic potential profiles during the *J-V* scans for the (**g**) top cathode and (**h**) bottom cathode device.

To investigate the processes underlying hysteresis we examined the evolution of open circuit photovoltage ($V_{oc}$) with steady state illumination, after preconditioning devices with a fixed bias voltage in the dark. This approach avoids the unnecessary complications introduced when analysing current-voltage sweeps, where both time and applied voltage are covariant. While monitoring the evolution of the $V_{oc}$ generated by the constant bias light (which we sometimes refer to as the background $V_{oc}$), a series of short (500 ns) laser pulses were simultaneously superimposed on the device to induce small perturbations in the photovoltage signal. Figure 2a shows a schematic of the experimental timeline for these 'transients of the transient' measurements. Analysis of the transient photovoltage perturbations gives information about changes in the movement and recombination kinetics of photogenerated charges as the background $V_{oc}$ evolves.

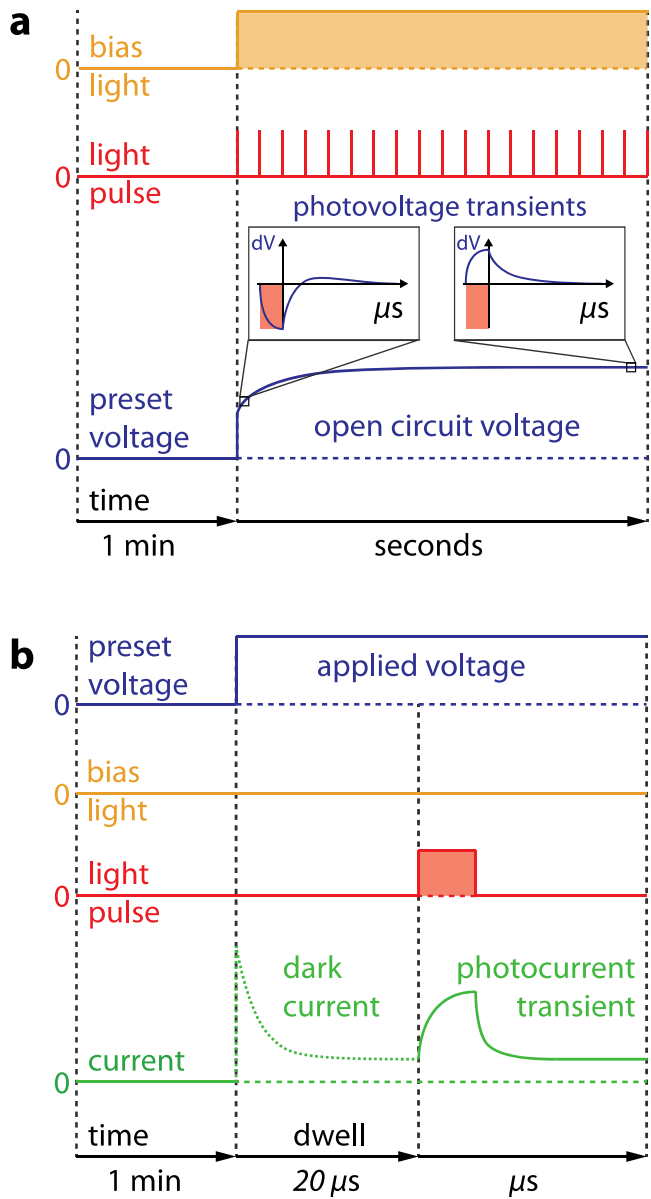

**Figure 2.** Experimental timelines for the optoelectronic transient measurements. (**a**) 'Transients of the transient' photovoltage measurement. The device is held in the dark at a preset voltage for 1 min before being switched to open circuit with 1 sun equivalent illumination. During the $V_{oc}$ evolution the cell is pulsed with a 638 nm laser at 1 second intervals and the resulting photovoltage transients are recorded. (**b**) 'Step-to-voltage' photocurrent transient measurement. The device is held in the dark for 1 min at a preset voltage (0 V in this study); then switched to an applied measurement voltage at which it is held for a further 20 μs, allowing the dark current to stabilise. A 10 μs near-infrared LED pulse is then used to excite the cell and the transient photocurrent is recorded.

Hybrid perovskite solar cells are often preconditioned using an applied forward bias or illuminated open circuit conditions prior to measurement. This procedure changes the polarization of the device to a state in which higher efficiency values can sometimes be inferred from *J-V* measurements than compared to short circuit or reverse bias preconditioning.[1,3] To explore this effect we have used two preconditions in this study: short circuit dark conditions ($V_{preset}$ = 0 V) where the device is polarized by the built-in potential between the contacts ($V_{bi}$ ~ 09 – 1.3)[41-43], or an applied forward bias ($V_{preset}$ = 1 or 1.2 V), which significantly reduces the potential, and thus the device polarisation, between contacts. These two states form the starting conditions for the subsequent transient measurements.

Figure 3a shows the evolution of photovoltage of a top cathode solar cell, which had been preconditioned with a forward bias of +1 V in the dark for one minute prior to switching the cell to open circuit and simultaneously turning the bias light on. After the initial development of the open circuit photovoltage to about 1 V (in < 50 µs), there is then a small increase in $V_{oc}$ (~10 mV) over the course of the measurement. Throughout the measurement there was no significant change in the shape and time constants of the transient photovoltage (figure 3b and inset in figure 3a). This is unsurprising since there is only a small change in the background $V_{oc}$ over the course of the measurement. Similarly stable results were also observed with this cell when it was preconditioned at short circuit ($V_{preset}$ = 0 V) in the dark instead of +1 V (see figure S2). These observations are consistent with the absence of significant hysteresis seen in the *J-V* curves in figure 1a.

In contrast, the bottom cathode devices (figure 1b and 1d) exhibited very different behaviour upon preconditioning at different bias voltages. Figures 3c and 3d show the $V_{oc}$ 'transients of the transient' measurement performed on a bottom cathode solar cell, which showed significant hysteresis (figure 1c). Following a forward bias (+1V) preconditioning step in the dark, the $V_{oc}$ declines steadily from around 850 mV to 720 mV in around 40 s under continuous illumination. Although the magnitudes of the small perturbation transient photovoltage decays decreased (figure 3d), the inset in figure 3c shows that the time constants (fitted using a double exponential function) during the measurement *increased* by a factor of approximately two over this time.

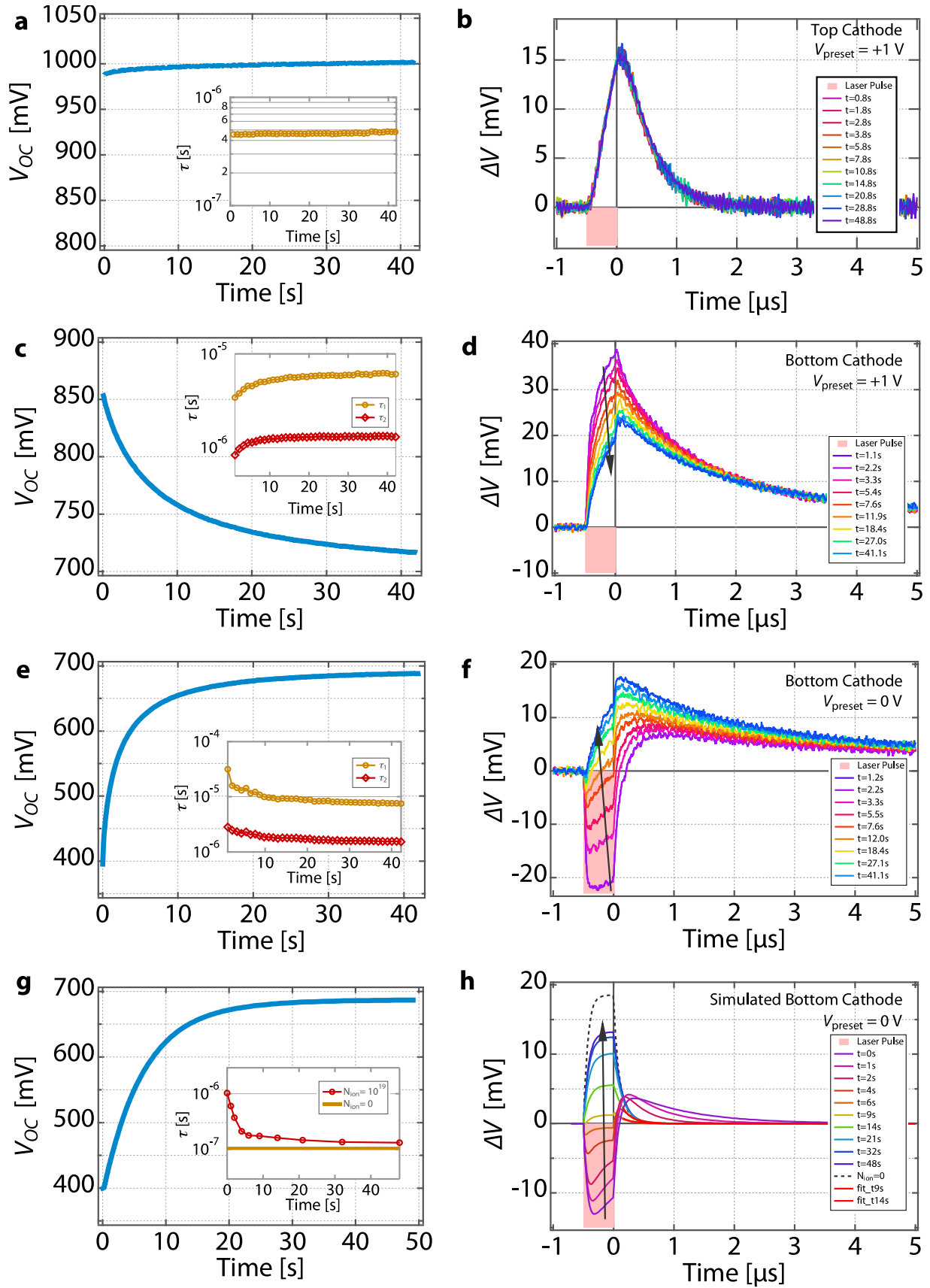

**Figure 3. Transient measurements of the open circuit photovoltage evolution.** Slow timescale evolution of the open circuit voltage (left column) and corresponding small

perturbation photovoltage transients (right column). The measurements were made using the protocol shown in figure 2a. The insets show the fitted decay time constants corresponding to the tails of photovoltage transients. Top cathode device preconditioned in the dark at +1 V (**a** and **b**). Bottom cathode device preconditioned in the dark at +1 V (**c** and **d**) and 0 V (**e** and **f**). Corresponding simulation of the photovoltage evolution and transient photovoltage measurements of a p-i-n device with mobile ions and high recombination in the p- and n- type regions with preconditioning of 0 V (**g** and **h**), c.f. figure 1d.

A zero-dimensional kinetic model of the device suggests that the factor of 2 increase in $\tau$ would correspond to a 36 mV *increase* in the photovoltage if the band gap and recombination reaction order were constant (see Note S1 for details). This is clearly inconsistent with the observed *decrease* in $V_{oc}$, suggesting a more sophisticated model is required to describe device behaviour, as will be discussed later.

Figure 3e shows the evolution of the bottom cathode device photovoltage following preconditioning at short circuit (0 V) in the dark. In this case the photovoltage rises by around 300 mV from 400 to 700 mV over 42 seconds. The simultaneous transient photovoltage signals in figure 3f exhibit anomalous behaviour: the change in photovoltage during the light pulses at early times of the slow background $V_{oc}$ evolution is negative. By comparison, the tails of the transients are all positive, and the time constants for their decay towards quasi-equilibrium *decrease* by a factor of 2 – 4 over the course of the measurement. We would expect this change to correspond to a *decrease* in $V_{oc}$ of between 36 and 72 mV if all else were constant (Note S1); this is also clearly inconsistent with the observed slow *increase* in voltage.

The negative deflection of the transient photovoltage measurements during the 500 ns transient light pulses indicates the existence of a positive internal current within the device (note that we define the sign of normal photocurrent to be negative, cf. figure 1). This additional positive current and associated negative displacement voltage at open circuit during the light pulse is apparent for the first ~ 20 seconds of the bias light exposure at open circuit. This implies that, during this time, a significant region of the internal electric field, is opposite to that expected from the built-in potential and driving photogenerated carriers towards the wrong contacts. An accumulation of space charge must exist to generate this opposing **E**-field; below we show this is likely to be near the interfaces.

Taken at face value the results in figures 3a – 3f might suggest that ion migration, which could cause this charge accumulation, is present in the bottom cathode architecture but not the top cathode devices (these do not show the negative photovoltage transients). To examine whether this is the case we also performed photocurrent transient measurements on both device types after stepping from a preconditioning bias voltage (0 V) to an applied forward bias near $V_{oc}$ (but below the built in voltage) in the dark (see the experimental timeline in figure 2b). The dwell time was sufficiently short to allow the dark current to stabilise without significant ion migration occurring. This was verified by the observation that the photocurrent transients did not vary following dwell times of up to at least 500 µs. By stepping to an applied electrical bias near $V_{oc}$, instead of using a bias light to generate an open circuit photovoltage, the charge carrier transport direction can be probed without flooding the device with photogenerated charges. The sign of the transient photocurrent reflects the direction of the dominant electric field in the cells.

The results of the measurements on the bottom cathode device are given in figure 4a. The control photocurrent transient measurement made at short circuit shows a negative photocurrent deflection, as expected. When the cell was switched to forward bias immediately prior to the transient measurement, we observe that the photocurrent transient is *positive*, consistent with the negative photovoltage transients observed for the bottom cathode device in figure 3e, which result from a positive internal photocurrent in the device. This observation confirms that there is an accumulation of space charge in the bottom cathode device causing an **E**-field opposing the built in potential.

Remarkably, when the measurement was repeated on the top cathode device, the photocurrent transient was similarly positive when the cell was stepped to a forward bias of 0.9 V (see figure 4b). This appears inconsistent with the purely positive photovoltage transients observed in figure 3b. The measurement indicates that, when there is no bias light flooding the device with photogenerated charge carriers, an **E**-field opposing the built-in potential initially exists. This is strong experimental evidence that there is also an accumulation of slow moving space charge in the top cathode architecture devices, despite the absence of significant hysteresis in the *J-V* scan at room temperature.

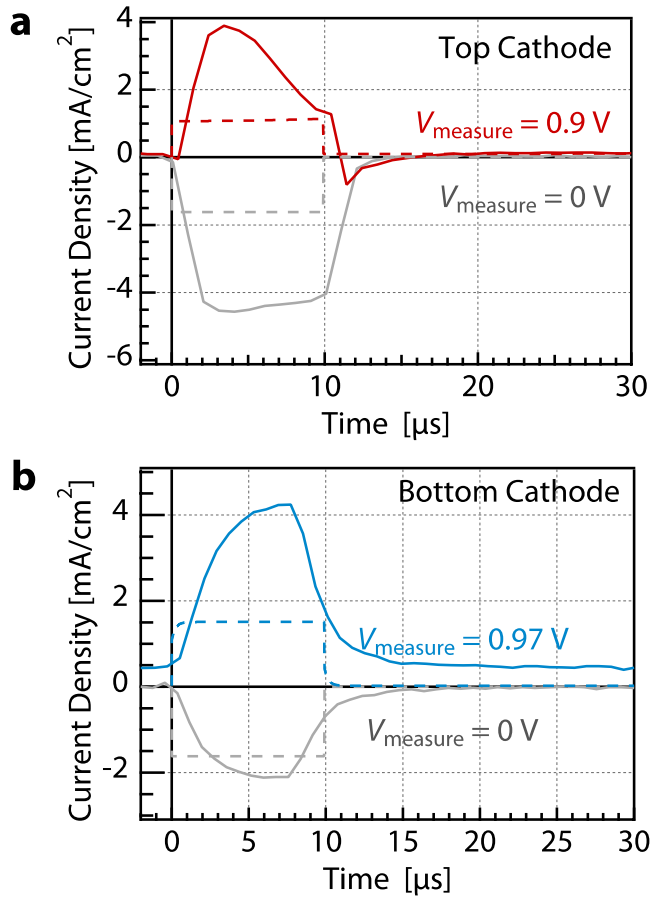

**Figure 4.** **Step-to-voltage transient photocurrent measurements.** The photocurrent transient measurements were taken using a 10 μs, 735 nm, pulse after a 20 μs dwell time following switching from 0 V to an applied forward bias in the dark ($V_{measure}$) following the protocol shown in figure 2b. (**a**) The red curve shows the positive photocurrent transient for a top cathode device with an applied measurement voltage of 0.9 V (**b**) The solid blue curve shows the positive transient from a bottom cathode device where 0.97 V in the dark was applied following preconditioning at 0 V in the dark. In both cases control photocurrent transients at short circuit are also presented (solid grey lines); these show negative photocurrents as expected. The dashed lines show the corresponding simulated photocurrent transients. Note that the rise and fall times of these transients are limited by the *RC* time constant of the device and the switch-on time of the LEDs used to create the light pulse.

## Simulation

As discussed in the introduction mobile ionic defects are widely thought to be intrinsically present in $CH_3NH_3PbI_3$. The accumulation of these charged defects at the contacts of devices due to the internal electric field within the perovskite layer has been used as a model to understand some hysteresis behaviour.[1,6,7,16,18] To test whether ion migration could explain the anomalous transient results observed here, we used a one-dimensional time-dependent drift-diffusion model that included a single, non-ionising mobile ionic species, electrons and holes (see Methods and Note S3 for details). The focus of this study was not to realistically simulate all aspects of the devices, but to explore principal device behaviour.

For simplicity we simulated the devices as p-i-n structures in which the intrinsic perovskite layer is sandwiched by contacts approximated by p-type and n-type regions with identical band gaps. The perovskite layer was set to contain a mean concentration of $10^{19}$ $cm^3$ positively charged mobile ionic species (these could correspond to $I^-$ vacancies for example) with a corresponding uniform concentration of negative static ionic species. The top cathode architecture was assumed to have no recombination in the perovskite/contact interfacial regions. The only difference in simulating the bottom cathode device was the introduction of Shockley Read Hall (SRH) recombination in the contact materials, to simulate recombination in these regions.

Figures 5a and 5b show the simulated energy level profiles and charge carrier density distributions of a top cathode device after reaching equilibrium at short circuit in the dark. The simulated data for the bottom cathode device under these conditions is virtually identical (data not shown). At equilibrium mobile ionic charge screens the cell's built in potential, created by the difference in Fermi energies of the p- and n-type 'contact' regions (grey shaded regions in figure 5). This results in an accumulation of space charge at the perovskite/contact interfaces with an associated strong electric field, and a field-free region in the bulk of the device. The details of this distribution depend on the defect concentration, built-in potential and dielectric constant.[18]

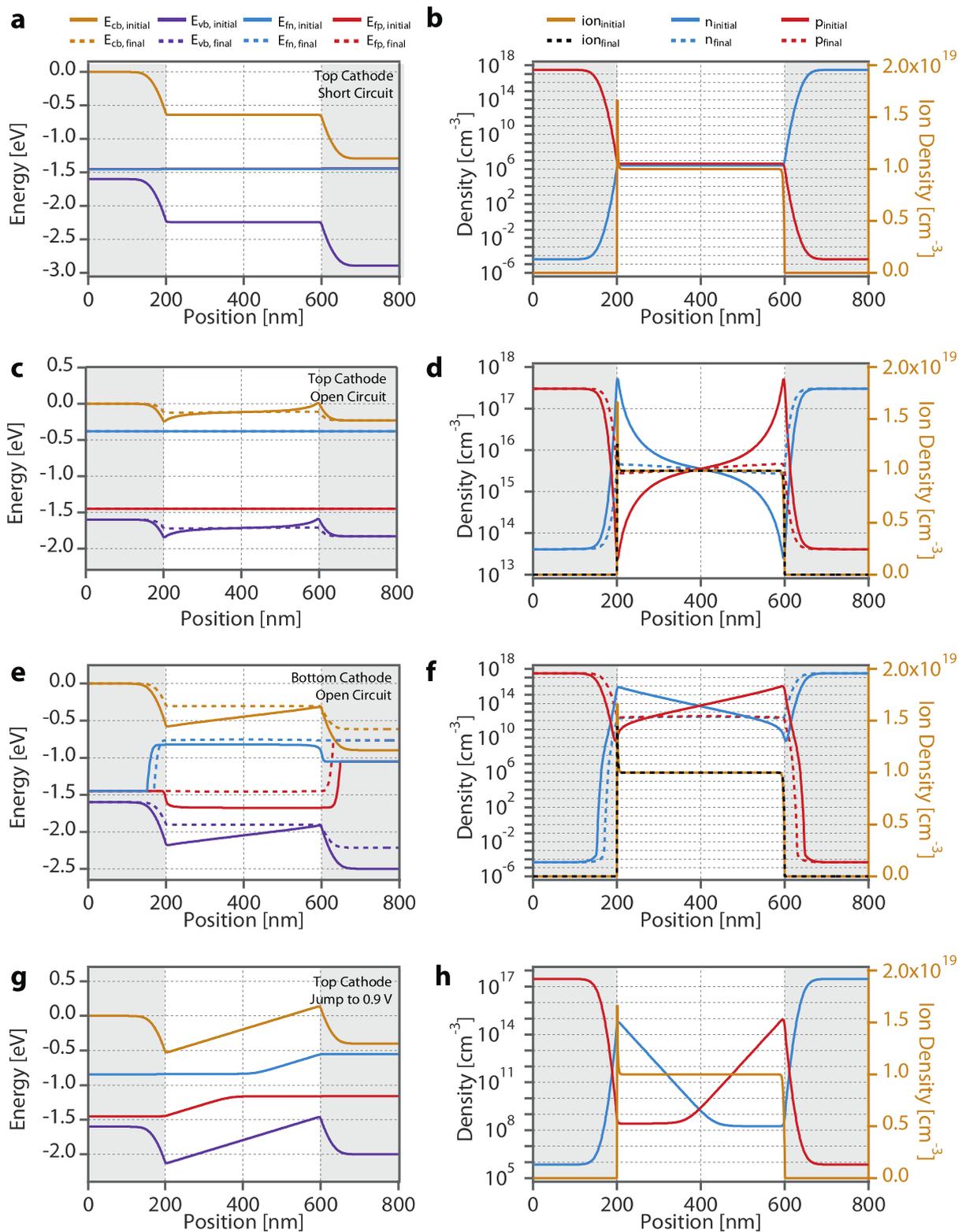

**Figure 5.** **Simulated energy level and charge density profiles.** (**a**) and (**b**) Top cathode cell at short circuit in the dark (virtually identical results were simulated for the bottom cathode device, data not shown). Mobile ionic charge has drifted to completely screen the built-in potential between the n-type and p-type contact materials. (**c**) and (**d**) Top cathode cell at open circuit under illumination following short circuit in the dark. (**e**) and (**f**) Bottom cathode cell at open circuit under illumination after short circuit in the dark. (**g**) and (**h**) Top cathode cell after stepping to an applied forward bias of +0.9 V from short circuit in the dark. In (**c**) – (**f**) initial states (shown after 50 µs) are indicated by solid lines while dashed lines designate final states (after 50 seconds of ion migration) under illumination at open circuit. The grey shaded regions represent the p- and n-type contact regions sandwiching the bulk perovskite layer. Note that the energy scale is referenced to the potential of the conduction band at the left-hand boundary on the semiconductor energy scale to give the potential energy of an electron (the electrochemical scale has the opposite sign). The meaning of the legend symbols is explicitly defined in Note S3.

When the top cathode device is then illuminated at open circuit (simulating the experimental conditions in figure S2), the initial build-up of photogenerated electrons and holes results in quasi-Fermi level splitting and the development of an open circuit potential, as expected. Immediately after illumination, the distribution of mobile ionic charges is as shown in figure 5b since they have not had time to move in response to the changed electric fields. In the presence of a photovoltage, this ionic charge distribution results in spatial electrostatic potential minima at the perovskite/contact interfaces, which we will refer to as 'valleys'. However, in the top cathode device, high concentrations of photogenerated electrons and holes rapidly redistribute to fill these valleys, resulting in screening of the initial **E**-field induced by ion accumulation (see figures 5c and 5d). Over a period of tens of seconds ions migrate away from the p-type region. This migration is a result of both the high ion concentration gradient at the interfaces and the reversal of the field direction, which now drives ions in the opposite direction to that during equilibration at short circuit (see figure S3a). There is an accompanying redistribution of the electrons and holes throughout this time. The key observation is that despite this ionic and electronic charge rearrangement, the change in $V_{oc}$ is very small (~ 1 mV, see figure S4), in agreement with the magnitude of the measured change seen in figures 3a and S2.

Consistent with observation, hysteresis is not present in the simulated *J-V* curves of the top cathode device (figure 1e) despite the reversed **E**-fields in the perovskite region during the forward scan (figures 1g and S5a). The absence of recombination in the interfacial regions allows the accumulation of photogenerated charge carriers in the valleys which partially screen the ionic charge. This allows the efficient collection of carriers by diffusion as indicated by the gradients of the electron quasi-Fermi levels shown in figure S5c. The simulated transients and *J-V* curves confirm that, despite the slow redistribution of ionic defects within the material, significant hysteresis is not expected in the absence of surface or interfacial recombination.

When recombination in the contact layers is included, the simulation results replicate the slow evolution of the $V_{oc}$ following dark short circuit preconditioning seen in bottom cathode devices (figures 3e and 3g). From the simulated energy and charge distribution diagrams shown in figures 5e, 5f and S3b (detail), it is apparent that potential valleys are also formed immediately following illumination. In this case however, photogenerated electrons and holes collected in the valleys rapidly recombine due to high rates of recombination in the n and p-type contact regions, and the electric field associated with ionic charge remains unscreened. Since the presence of the photovoltage partially negates the built-in potential, the concentration of ionic defects at the contacts decreases as the defects migrate away, until the **E**-field in the bulk of the cell is once again zero, consistent with recent scanning Kelvin probe observations[44]. At this point, the $V_{oc}$ reaches a plateau. The evolution of $V_{oc}$ during this time is dictated by 1. ionic migration, 2. electronic charge rearrangement in response to ionic migration, and 3. an increase in the concentration of photogenerated charge carriers due to redistribution away from fast SRH recombination centres in the contacts. An opposite process occurs for a bottom cathode device relaxing from forward bias preconditioning (figure S6). These processes are also responsible for the difference in $V_{oc}$ between forward and reverse scans in the *J-V* curve (see figure 1f). Figures 1g, 1h, S5a and S5b show that the potential profiles are similar for both bottom cathode and top cathode simulations at low forward bias. However recombination of carriers driven to the interfaces by the reversed **E**-field during the forward scan results in a significant loss of photocurrent (figure S5d). The valleys are not present for the reverse scan so charge collection is more efficient. The simulated *J-Vs* showed similar hysteresis indices (HIs) for the same scan speed and similar scan protocol (see Methods), with HIs of 0.00 and 1.84 compared to measured values measured for these devices of 0.05 and 1.71 for the top cathode and bottom cathode architectures respectively. We note that the degree of hysteresis observed in a *J-V* measurement is sensitive to the voltage scan rate,[1] since this determines whether mobile ionic defects have time to react to the changing applied potential as has recently been simulated.[18] If the ratio of the scan rate to ionic mobility is very large the ions will appear 'frozen' in place during the scan, alternatively if the ratio goes to zero then ions will reach an equilibrium distribution for each voltage in both scan directions, and no hysteresis will be seen in either case.

Our simulations also reproduce the anomalous transient photovoltage behaviour observed at early times during the $V_{oc}$ evolution (figures 3f and 3h, and figure S4 for the dark +1 V preconditioning). The negative transients are explained by the drift of the additional photogenerated charges to the valleys during the laser pulse (a positive internal photocurrent). Once the light pulse ends, the associated positive internal photocurrent stops and the excess electrons and holes then contribute to a positive deflection of the transient photovoltage in the normal way. This excess then decays away on a timescale reflecting the recombination kinetics of the device in its current state. The details underlying the photovoltage decay time constants are considerably more complex than the zero-dimensional model mentioned above (Note S1), owing to the spatial separation of electrons, holes and recombination regions. The simulation results in transient photovoltage decay time constants that decrease as the $V_{oc}$ increases, consistent with experimental trend (see figure 3e and 3g insets).

Simulation of the step-to-voltage transient photocurrent measurements are shown in figure 4, and figure 5g and 5h. The presence or absence of interfacial recombination makes almost no difference to the simulated results since there are no photogenerated charge carriers prior to the pulse. Due to the ionic charge distribution, positive photocurrent transients following the step to a forward bias measurement voltage are observed in the simulations for both the top cathode and bottom cathode architectures, consistent with measured positive photocurrent transients in both cell types.

The simulations presented here contain only two features that differ from an ideal p-i-n solar cell: mobile ionic defects, which can result in a slow redistribution of charge in the perovskite layer; and for the bottom cathode cell, high rates of recombination in the p-type and n-type contact layers. These additions were sufficient to explain key features of the complex 'anomalous' behaviour observed in our experiments on timescales ranging from $10^{-8} - 10^2$ s with reasonable quantitative agreement.

For completeness, simulations with zero and low concentrations of mobile ions, with and without interfacial recombination, were tested. The results showed that mobile ion concentrations $> 10^{17}$ cm$^{-3}$ are required to reproduce the observed behaviour (figure S7).

Other possible recombination schemes in the presence of ionic migration were also investigated (Note S2). These indicated that: the recombination type in the contact regions did not need to be specifically SRH to reproduce the experimental results; and that high band-to-band recombination rates throughout the device or in the perovskite layer alone cannot reproduce the results (figure S8). Simulation of significant recombination in only one contact layer yielded interesting results: it reproduced negative photovoltage transients at early times, but the $V_{oc}$ showed an initial increase which peaked after about 5 seconds followed by a decline to a plateau (figure S8). This can be rationalised when one considers that the 'valley' at the p-type contact will initially accumulate electrons without significant surface recombination but will eventually 'discharge' these for recombination at the n-type contact as the ions redistribute (see figure S9). We have frequently observed biphasic evolution of the photovoltage with time in devices (e.g. figure S10); this could be consistent with an asymmetry in the recombination rates in the two contact materials, where the single sided recombination is an extreme case of this possibility.

In summary, our study confirms that $CH_3NH_3PbI_3$ shows behaviour consistent with a mixed ionic/electronic conductor at room temperature. We have used optoelectronic photocurrent transient measurements with no bias light to demonstrate that the effects of ionic migration can be observed in devices that exhibit 'hysteresis free' *J-V* and transient $V_{oc}$ behaviour as well as those that show hysteresis. Simulation of the measurements shows that *J-V* hysteresis, slow timescale evolution of the $V_{oc}$ and negative photovoltage transient behaviour are reproduced by the combination of ionic migration *with* high rates of recombination in the perovskite/contact interfaces. We note that this could also include recombination via pin holes, and could be partially attributable to the different substrates, morphology and perovskite deposition techniques used in processing each architecture.[29] Recent studies using alloyed hybrid perovskite preparations sometimes show relatively minimal hysteresis before aging.[45,46] However, as well as 'standard' hysteresis,[47,48] in some cases inverted hysteresis is observed in these materials.[39] This suggests recombination may be mediated by mobile defects themselves in addition to the interfaces which is likely to be an interesting area for future investigation.

Our results provide experimental confirmation of the predictions from simulations by van Reenen et al.[16] that both ionic migration and interfacial recombination are required for hysteresis to be observed. We also conclude that PCBM does indeed passivate interfacial recombination at the cathode relative to metal oxide contact materials, as suggested in previous studies.[30,36,38] However, our observations are not consistent with the hypothesis that PCBM reduces hysteresis by preventing the diffusion of ionic defects along grain boundaries.[30] In addition to demonstrating the role of interfacial recombination in the presence of mobile ions in a semiconductor, our study demonstrates the viability of controlling the measureable consequences of this ionic migration. This suggests the interesting possibility of exploiting these effects for other electronic applications where a memory of previous operating conditions would influence device behaviour.[14,15]

## Methods

### Devices

The planar bottom cathode devices had the following stack of layers: FTO glass/dense-TiO2 (~50nm)/$CH_3NH_3PbI_3$ (~300nm)/Spiro-OMeTAD(~200nm)/Au (80nm) with an active area of 0.08 cm$^2$; the cells were prepared as described in reference [3]. The planar top cathode stack of layers was: ITO glass/PEDOT:PSS (30nm)/CH3NH3PbI3 (300nm)/PCBM (85nm)/Ca (20nm)/Al (100nm) with an active area of 0.1 cm$^2$; the cells were prepared as described in reference [34]. The key transient behaviour presented in this study was reproduced in all ten working devices measured.

### Optoelectronic characterization

Current-voltage sweeps of the devices were made using a Keithley 236 Source Measure Unit and a xenon lamp solar simulator with AM1.5G filters (Oriel Instruments). The illumination intensity was adjusted to be equivalent to 100 mW/cm² using a using a calibrated filtered Si photodiode (Osram BPW21). The *J-V* measurement protocol was as follows for both the light and the dark measurements. The cell was left at -1 V in the dark for approximately 30 s. In the case of the light measurements the solar simulator shutter was then opened. The applied voltage was then swept from -1 V to +1.2 V at a rate of approximately 40 mV s$^{-1}$ (forward scan) and the current density measured, the optical shutter was then closed. The cells were then held at +1.2 V for a few seconds before the shutter was opened and the voltage swept back to -1 V at 40 mV s$^{-1}$ (reverse scan). As means to compare the degree of hysteresis between devices we define a hysteresis index (HI = [$P_{max,r}/P_{max,f}$] - 1) for a given scan rate in terms of the maximum power points on the reverse scan, $P_{max, r}$, and the forward scan, $P_{max, f}$. This differs slightly from the hysteresis index introduced by Kim and Park.[49] Top cathode devices showed minor variation in hysteresis (mean HI = 0.01 ± 0.03 from 7 devices), while the bottom cathode devices showed greater hysteresis with more variation (mean HI = 4.5 ± 5.5 from 3 devices with minimum HI = 0.87).

Transient of the transient photovoltage measurements were made using a National Instruments USB-6361 data acquisition card to monitor the slow voltage transients (generated by a white bias light) and a Tektronix DPO5104B digital oscilloscope to monitor the fast voltage transients (generated by a red laser pulse). The two measurements were performed simultaneously (see Figure 2a for the experimental timeline). The fast voltage transients were collected every second for approximately 45 seconds, averaging approximately 20 curves over approximately 200 ms. The white bias light was provided by an array of cool-white LEDs (Luxeonstar), calibrated to 1 sun equivalent with a silicon photodiode. The 500 ns laser pulse was provided by a digitally-modulated Omicron PhoxX+ 638 nm diode laser, with 100 Hz repetition rate. The laser spot size was expanded to cover the active pixel and the continuous wave intensity over the cell pixel area was approximately 550 mW cm$^{-2}$ during the pulse. The preconditioning bias was applied using the data acquisition card. The system was controlled by a custom Labview code. The measurement sequence is shown in figure 2a.

Single exponential small pertubation photovoltage decays (Δ$V$) were seen in some 'top cathode' devices, consistent with organic and dye sensitised solar cells. However for many devices, most notably those with the 'bottom cathode' architecture, the photovoltage transient decays could only be accurately fit using a bi-exponential function as has been reported previously:[3,4,50,51]

$$\Delta V = A_1 e^{-\frac{t}{\tau_1}} + A_2 e^{-\frac{t}{\tau_2}}, \tag{6}$$

where $A_1$ and $A_2$ are the amplitudes of the two components and $t$ is time. We have not yet confidently assigned the decay constants, $\tau_1$ and $\tau_2$ to specific physical phenomena although our observations suggests the biphasic decay is related to hysteresis and the probable asymmetry of recombination rates at the contacts (see figures S8 and S9). We note that the constants appear to vary proportionately and thus we use them to parameterise relative changes in the recombination lifetime.

The step-to-voltage photocurrent transients were made using light pulses from a 735 nm LED (Ledengin LZ1-00R300). The turn off time of the 735 nm LED was ≤ 200 ns as measured by a fast silicon diode and a GHz oscilloscope. Turn on time for the LED was ≤ 3 μs. The pulse intensity gave an absorbed photon flux approximately equivalent to 0.5 suns. Applied potential was supplied by a National Instrument USB-6251 multi-function DAQ board. Claimed output voltage slew time is 20V/μs and settling time <1 μs. Current was measured on the same USB-6251 with 16 bit resolution and 0.8μs per point.

**Drift-diffusion model**

A 1-dimensional drift diffusion model was implemented to simulate the results using MATLAB's built-in partial differential equation solver for parabolic and elliptic equations (pdepe). The full details can be found in Note S3. The code solves the continuity and Poisson's equations (equations S16 – S22) for electrons, holes and positively charged mobile ionic defects (confined to the perovskite layer) and the electrostatic potential as a function of space and time. Note that neither the mobile nor the static ionic defects were modelled to induce doping effects which could liberate free electrons and holes in the conduction and valence bands, so the only role of the mobile defects in the simulation is to allow the distribution of charge in the device to change. Equilibrium and doping electron and hole densities were calculated using Boltzmann statistics (equations S23 – S25).

To simulate current-voltage scans and current transients, a p-type/intrinsic/n-type (p-i-n) structure device (figure S11a) was used with an increasing or decreasing series of fixed potential difference boundary conditions (equations S29 – S34) for each time step. The scan protocol used in the *J-V* simulations was similar to that used experimentally (with a scan rate of 40 mV s$^{-1}$), the final state of the device after the forward scan was used as the starting condition for reverse scan.

Conventionally the open circuit voltage is found by using Newton's method to solve for zero current at the boundaries. In order to accelerate calculation times and enable direct readout of the open circuit voltage, we used the method of image charges and devised a symmetric p-i-n (s-p-i-n) cell (figure S11b). This approach enables simple Dirichlet boundary conditions to be employed by setting the potential at both boundaries equal to zero (equations S35 - S40). Since charge densities in the two cells are a mirror of one another, the current and electric field at the mid-point of the device go to zero. However, the potential at the mid-point is free to change in response to changes in carrier concentration profiles in accordance with Poisson's equation. Provided that the depletion widths at the intrinsic/doped layer interfaces are significantly less than the layer thicknesses themselves, the doping concentrations in the n- and p-type layers generate the built-in voltage in the device. The $V_{oc}$ of the cell is obtained by taking the difference in electron quasi-Fermi energy, $E_{Fn}$ at the mid-point of the device (cathode) and the hole quasi-Fermi energy, $E_{Fp}$, at the left-hand boundary (anode) (equation S26). For clarity, band diagram and charge density figures included herein for the s-p-i-n model, only show the left-hand half of the device.

The model used a mesh with a linear grid $x$ spacing of 0.67 nm per point (1200 and 2400 points for p-i-n and s-p-i-n models respectively), which is marginally larger than a $CH_3NH_3PbI_3$ lattice cage width of 0.63 nm (ref. [52]). Examples of data calculated with other grid spacings are shown in figure S12; we chose 0.67 nm as a compromise between numerical accuracy and calculation time.

In order to further accelerate calculation times, generation in the active layer was uniform, and adjusted to yield an absorbed photon flux equivalent to 16 mA cm$^{-2}$ ($2.5 \times 10^{21}$ cm$^{-3}$s$^{-1}$). Voltage transients and current transients were taken using fluxes of 16 mA cm$^{-2}$ and 3.2 mA cm$^{-2}$ respectively. In simulating the top cathode device, SRH recombination was switched off, while for the bottom cathode devices, an SRH recombination was implemented in the contact regions with a time constant of $2 \times 10^{-15}$ s. The band-to-band recombination coefficient was chosen ($10^{-10}$ cm$^{-3}$s$^{-1}$) such that the time constants for the $V_{oc}$ transients in the top cathode device were similar those observed experimentally (figure 3b). While unrealistic, including SRH recombination throughout the thickness of the contact regions rather than solely at the interfaces circumvented numerical inaccuracies resulting from the combination of a strong electric field and high recombination rates at the perovskite contact interfaces. Using this method approximately 75% of the total SRH recombination takes place within the first 10 nm of the contact region (e.g. between 190 nm and 200 nm in the p-type region), yet the same $V_{oc}$ was obtainable using an order of magnitude higher recombination time constant than when a 5 nm recombination layer was used at the interface.

In the bottom cathode device, both the SRH time constant and trap energies determine the overall rate of SRH recombination (equation S28). In the absence of data regarding possible trap energies, the levels were chosen arbitrarily to be shallow (0.2 eV below the conduction band and 0.2 eV above the valence band for the n and p-type regions respectively). Since deeper trap energies combined with a lower recombination time constant yield similar behaviour, our simulations do not provide an estimate for the SRH time constant in real devices. The combination of SRH trap energy and time constant were chosen to result in a steady state $V_{oc}$ value approximately corresponding to that observed in the bottom cathode device.

The timescales and magnitudes of the $V_{oc}$ transients were sensitive to choice of built-in voltage (which was not adjusted for the different architectures), SRH recombination rates, initial ionic charge density and ion mobility.

## Acknowledgements

We thank Diego Alonso Alverez and Andrew McMahon for consultation on the drift diffusion simulation, and Davide Moia, Joel Troughton and Matthew Carnie for helpful discussions on devices and measurements. We thank James Durrant for use of facilities and the WAG funded Sêr Cymru Solar project for funding. AMT thanks the Imperial College Junior Research Fellowship Scheme for support. The authors are grateful to the UK Engineering and Physical Sciences Research Council for financial support (grants EP/J002305/1, EP/M023532/1, EP/I019278/1, EP/M025020/1, EP/G037515/1 and EP/M014797/1).

**Author contributions**

PRFB and BOR designed the study. AMT, PC and BOR performed the experimental measurements. AMT and BOR designed and built the experimental setups. PC and PRFB performed the simulations. DB and LX fabricated the samples. All authors contributed to the interpretation of results and preparation of the manuscript.

**Additional information**

Supplementary information is available in the online version of the paper. Reprints and permissions information is available online at www.nature.com/reprints. Correspondence and requests for materials should be addressed to PRFB.

**Competing financial interests**

The authors declare no competing financial interests.

**Supplementary Materials**

**Evidence for ion migration in hybrid perovskite solar cells with minimal hysteresis**
*Philip Calado[1§], Andrew M. Telford[1§], Daniel Bryant[2,3], Xiaoe Li[3], Jenny Nelson[1,3], Brian C. O'Regan[4**], Piers R. F. Barnes[1*]*

1   Department of Physics, Imperial College London, SW7 2AZ, UK
2   Department of Chemistry, Imperial College London, SW7 2AZ, UK
3   SPECIFIC, Swansea University, SA12 7AX, UK
4   Sunlight Scientific, 1190 Oxford Street, Berkeley CA, 94707, USA

§   These two authors have contributed equally to this work
*   piers.barnes@imperial.ac.uk
**  bor@borski.demon.co.uk

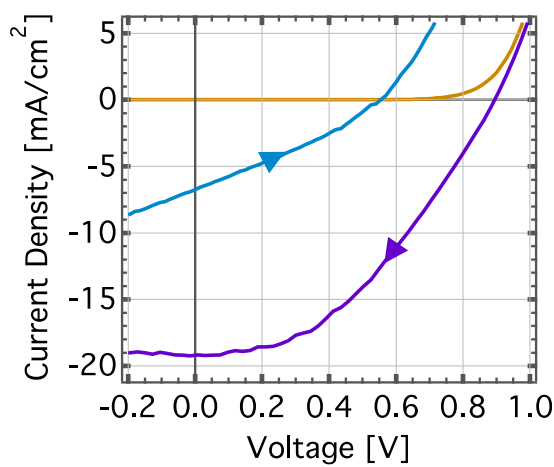

**Figure S1.     Hysteresis in a top cathode cell using a ZnO electron collection layer instead of PCBM.** Forward and reverse current-voltage scan (125 mV/s) for a ITO/PEDOT:PSS /CH$_3$NH$_3$PbI$_3$/ZnO/Al perovskite solar cell exhibiting large *J-V* hysteresis (ITO/PEDOT:PSS /CH$_3$NH$_3$PbI$_3$/PCBM/Al devices showed minimal hysteresis *cf.* figure 1c). The ITO/PEDOT:PSS /CH$_3$NH$_3$PbI$_3$ stack of layers was fabricated as described in reference [1]. The CH$_3$NH$_3$PbI$_3$ layer was then contacted with a layer of ZnO nanoparticles. The 20 mg/ml solution of ZnO nanoparticles was prepared as described in reference [2], only in this instance the solvent was replaced completely with IPA. The nanoparticle solution was spin cast on top of the perovskite layer at 2000 rpm for 45 seconds followed by heating at 120 °C for 5 minutes. An evaporated Al layer was then deposited in the same way as the PCBM containing device. A control ITO/PEDOT:PSS /CH$_3$NH$_3$PbI$_3$/PCBM/Al device was also prepared in which pure IPA was spun onto the PCBM layer and then heated prior to the evaporation of Al. The current-voltage sweeps for this control device were 'hysteresis

free', similar to those shown in figure 1c, and indicate that the IPA solvent and heating steps are not responsible for the presence of hysteresis in the device containing ZnO.

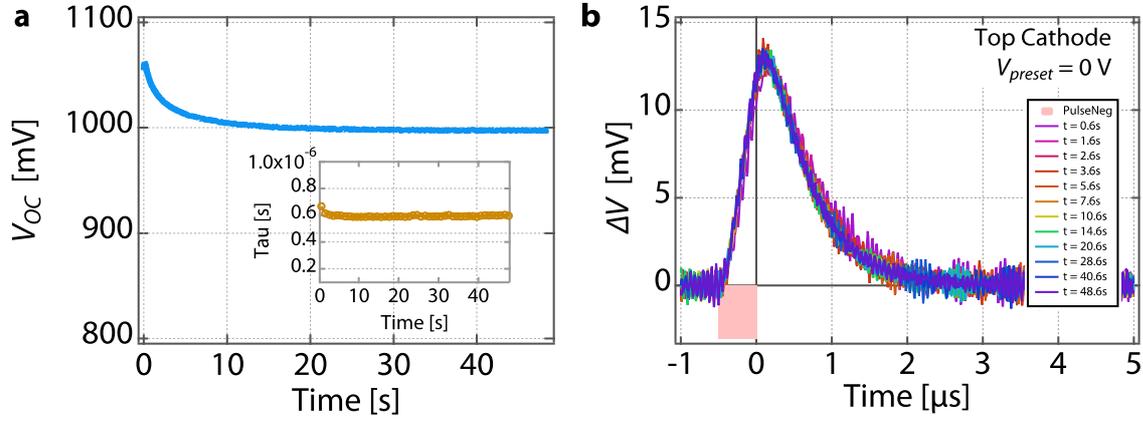

**Figure S2. Top cathode cell voltage transients of the transient measurements.**
Transients of the transient photovoltage measurement for the top cathode device with $V_{preset} = 0$ V. Both the open circuit voltage and photovoltage transients remain relatively stable throughout the measurement, similar to when the cell is preconditioned at forward bias cf. figures 3a and 3b.

**Note S1.  Derivation of zero dimensional kinetic model of recombination**

$CH_3NH_3PbI_3$ is assumed to be an intrinsic semiconductor, the electron and hole concentrations [cm$^{-3}$] are approximated by Boltzmann statistics and given by:

$$n = N_c \exp\frac{E_{Fn}-E_c}{k_B T} \quad \text{(S1)}$$

and

$$p = N_v \exp\frac{E_v-E_{Fp}}{k_B T} \quad \text{(S2)}$$

where $N_c$ and $N_v$ are the effective density of states at the conduction band and valence edges which have energies $E_c$ and $E_v$ respectively. $E_{Fn}$ and $E_{Fp}$ are the electron and hole quasi-Fermi energies, $k_B$ is Boltzmann's constant and $T$ is the temperature.

The mean effective recombination rate, $R$, in the device is approximated as a power law

$$R = k(np)^{\gamma/2} \quad \text{(S3)}$$

where $k$ is the effective recombination rate constant [cm$^{3(\gamma-1)}$ s$^{-1}$]. If the concentration of electrons is equal to the concentration of holes ($n = p$) then this expression becomes:

$$R = kn^\gamma \quad \text{(S4)}$$

where $\gamma$ is the recombination reaction order. For example, if simple band-to-band recombination (bimolecular) is dominant then $\gamma = 2$, if Shockley-Read-Hall recombination via mid-gap states dominates then $\gamma = 1$.

In this study we measured the time constant, $\tau$, of small perturbation transient decays. We wish to find the relationship between this measured values and the recombination rate constant, $k$. We follow the analysis in O'Regan et al.[3]: For a sufficiently small perturbation, the change in photovoltage is proportional to the change in carrier concentration in the device ($\Delta V_{oc} \propto \Delta n$). Consequently, the small perturbation photovoltage decay time constant, $\tau$, describes the rate at which the excess charge carriers decay:

$$\frac{d\Delta n}{dt} = -\frac{\Delta n}{\tau} \quad \text{(S5)}$$

This can be described in terms of the total recombination rate using equation S4:

$$-\frac{d\Delta n}{dt} = k(n + \Delta n)^\gamma - kn^\gamma \approx kn^\gamma\left(1 + \gamma\frac{\Delta n}{n}\right) - kn^\gamma = k\gamma n^{\gamma-1}\Delta n \quad \text{(S6)}$$

Using equation S5 we can then write equation S6 as:

$$k \approx \frac{n^{1-\gamma}}{\gamma\tau} \quad \text{(S7)}$$

Given that the cell runs at quasi-steady state conditions we may state that the recombination rate is equal to the generation rate, $G = R$, then substituting equation S7 into equation S3 we can write:

$$G \approx \frac{(np)^{(1-\gamma)/2}}{\gamma\tau}(np)^{\gamma/2} = \frac{(np)^{1/2}}{\gamma\tau} \quad \text{(S8)}$$

Substituting the expressions S1 and S2 into S8 gives:

$$G \approx \frac{(N_c N_v)^{1/2}\exp\left[\frac{E_{Fn}-E_{Fp}+E_v-E_c}{2k_BT}\right]}{\gamma\tau}$$

This can be rearranged to give an expression for the $V_{oc}$ by finding the difference between $E_{Fn}$ and $E_{Fp}$:

$$qV_{oc} \approx E_c - E_v + 2k_BT\log(\gamma\tau G) - k_BT\log(N_cN_v), \tag{S9}$$

where in principle the only variables are $\gamma$, $\tau$ and $G$. During the transients of the transient experiment the bias light, and thus $G$ is constant. If $\gamma$ is also constant and the model were valid then the change in $V_{oc}$ during the experiment would be given $2k_BT\log(\tau)$.

**Supplementary simulation data**

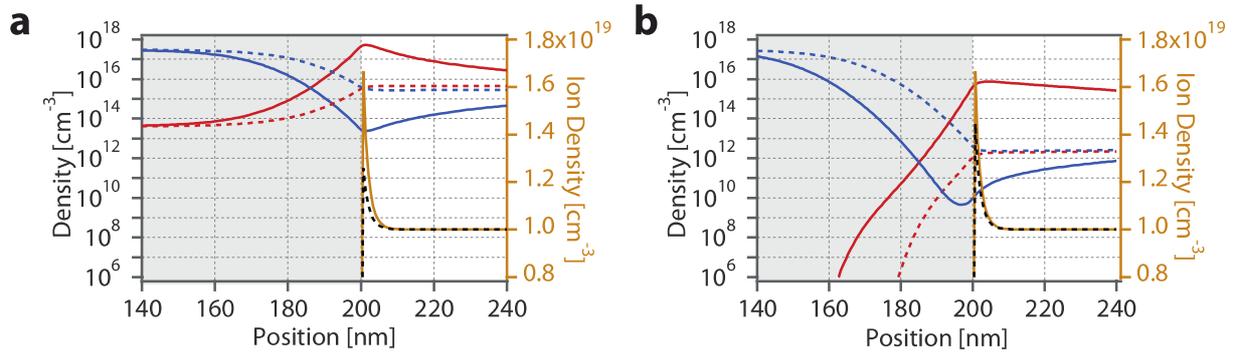

**Figure S3.** **Details of simulated charge density at interface.** Details from the device simulation charge densities for (**a**) top cathode and (**b**) bottom cathode devices showing the p-type/perovskite interface, change in ionic charge density and the depletion regions.

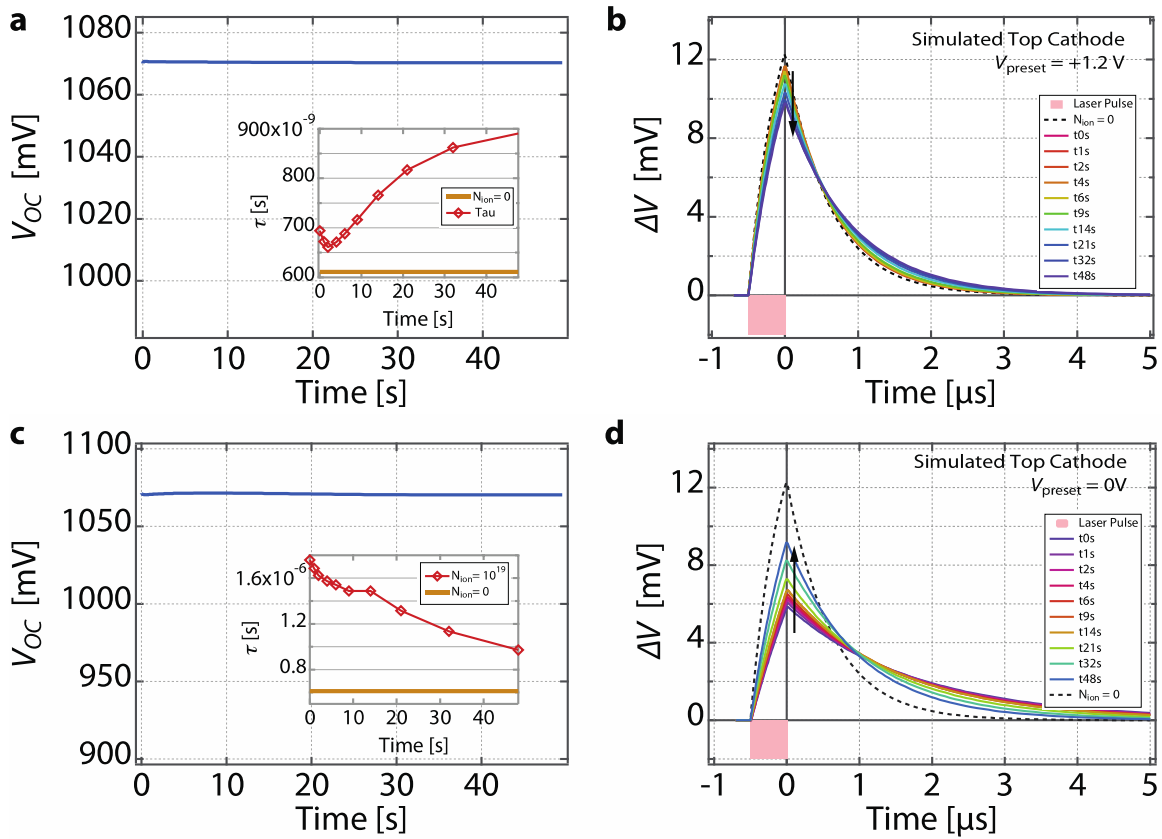

**Figure S4.** **Top cathode cell voltage transients of the transient simulations.** Simulated transients of the transient photovoltage measurement for a device with the SRH recombination coefficient in the p and n-type contact layers set to zero with (**a**) $V_{preset}$ = +1.2 V and (**b**) $V_{preset}$ = 0 V. A forward bias value of greater than the $V_{oc}$ of the cell was used as in the experiment cf. figures 1c, 1e and 3b. The small change in TPV magnitude observed in the simulated device could be attributed to the simplicity of using a p-i-n architecture and the absence of minority carrier accumulation at interfaces due to blocking contacts. An incorrect built-in voltage would also lead to the possibility of electronic charge not being able to fully screen the electric field created by ionic displacement after the cell has been switched to open circuit.

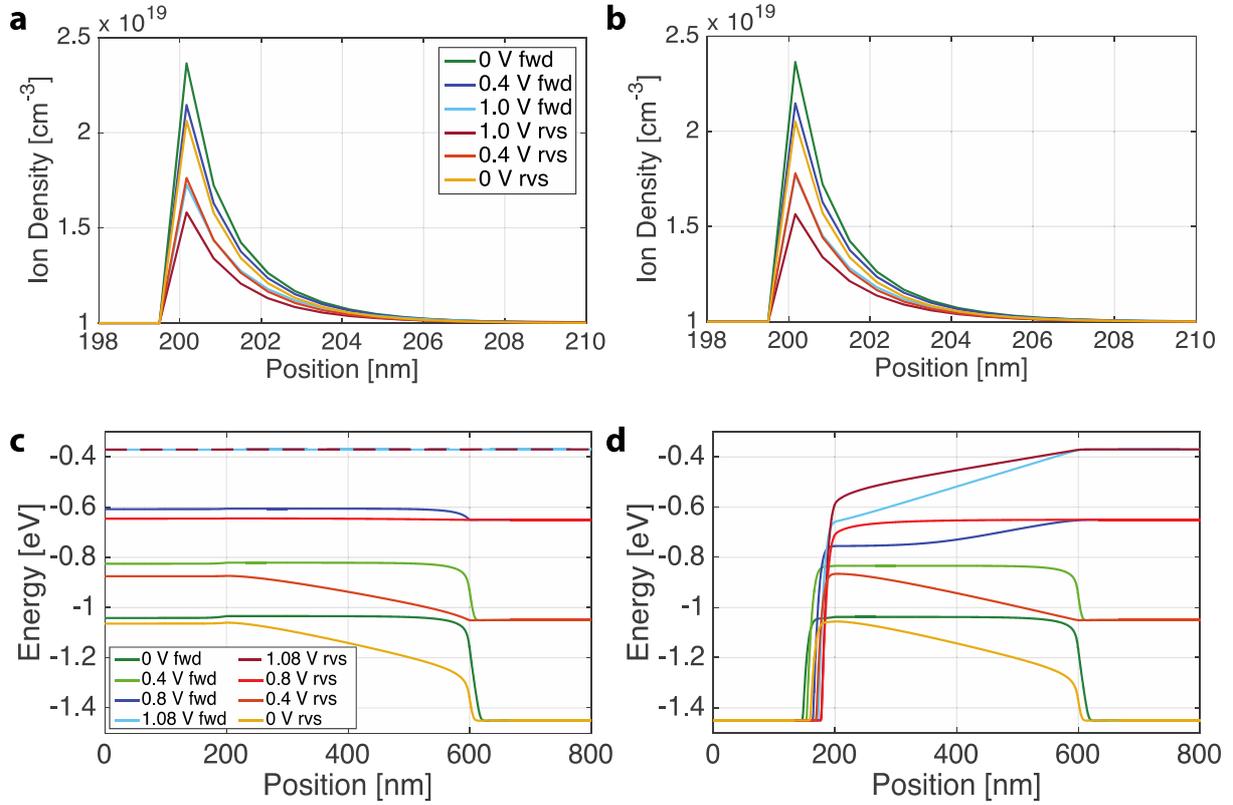

**Figure S5. Simulated ionic charge distributions and quasi-Fermi levels during *J-V* scans shown in figure 1.** The distribution of mobile ionic charge is shown near the p-type/perovskite interface (at 200 nm) for the (**a**) top cathode device and (**b**) bottom cathode device. The corresponding electron quasi-Fermi levels are shown for the complete devices in (**c**) and (**d**). The simulation outputs are shown for different applied voltage points (0, 0.4 and 1 V) during the *J-V* scans at 40 mV s$^{-1}$ from reverse to forward bias (fwd) and from forward to reverse bias (rvs).

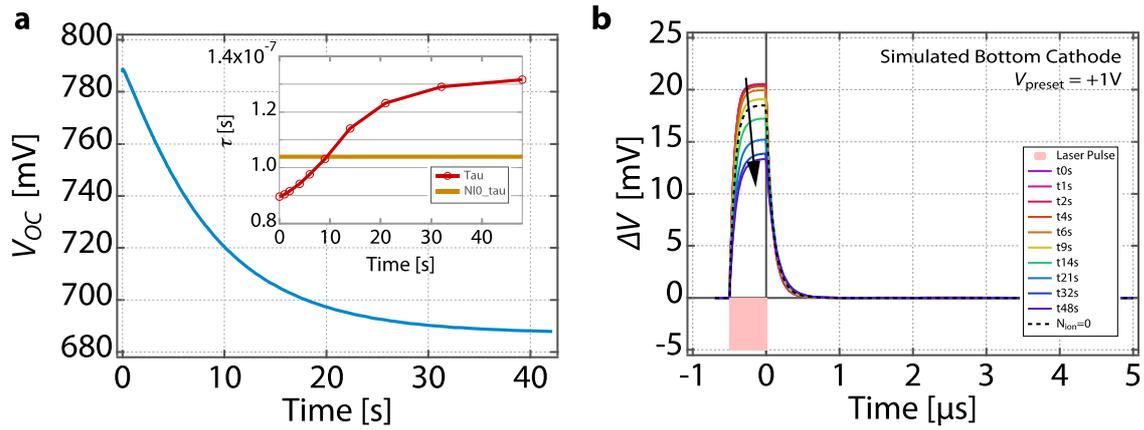

**Figure S6.** **Bottom cathode cell voltage transients of the transient simulations with forward bias preconditioning.** Simulated transient of the transient photovoltage measurement for a bottom cathode device ($\tau_{n,SRH} = \tau_{p,SRH} = 2 \times 10^{-15}$ s) after preconditioning with +1 V forward bias. See figures 3c and 3d of the main text for experimental equivalent. We note that the decay time constant after preconditioning with +1 V in the simulation is 1 – 2 orders of magnitude lower than observed experimentally. This could be attributed to an unrealistically high rate constant for band-to-band recombination or the absence of electron and hole blocking layers in the simulation, since no adjustment for possible variation in the built-in potential has been made.

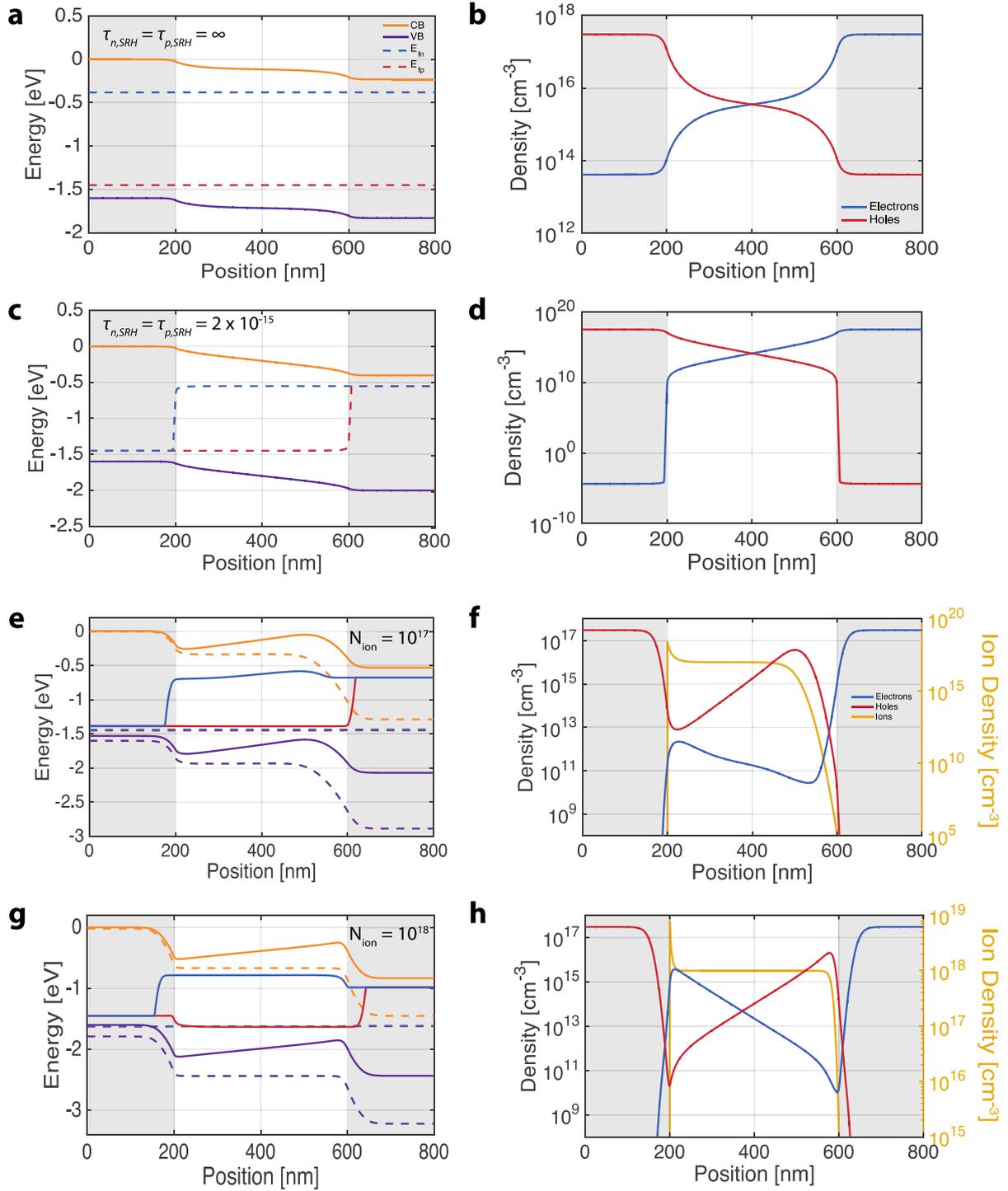

**Figure S7. Simulated effects of varying ion defect density and contact recombination.** Band diagrams (left) and charge densities (right) for simulated p-i-n structure solar cells at open circuit with a uniform carrier generation profile of $2.5\times10^{21}$ cm$^{-3}$s$^{-1}$. (**a**) and (**b**) Top cathode cell without mobile ion defects ($N_{ion} = 0$ cm$^{-3}$) or SRH recombination in the contacts ($\tau_{n,SRH} = \tau_{p,SRH} = \infty$ s). Standard quasi-Fermi level splitting is observed. (**c**) and (**d**) Bottom cathode cell without mobile ion defects ($N_{ion} = 0$ cm$^{-3}$) and with a high rate of SRH recombination ($\tau_{n,SRH} = \tau_{p,SRH} = 2\times10^{-15}$ s) in the contact regions. (**e**) and (**f**) Bottom cathode cell ($\tau_{n,SRH} = \tau_{p,SRH} = 2\times10^{-15}$ s) with $10^{17}$ cm$^{-3}$ mobile ionic defects after the cell has equilibrated at short circuit (dashed

lines) and been switched to open circuit (solid lines) with generation on. While the ion density is large enough to screen out the field between approximately $x = 210$ nm and $x = 440$ nm, the asymmetry caused by a single ionic carrier type and associated giant Debye length at the perovskite/n-type interface results in the formation of a valley further from the contact. Although faster recombination is switched on in the contacts, a greater proportion of recombination takes place in the bulk compared to the cases in which $N_{ion} = 10^{18}$ cm$^{-3}$ and $N_{ion} = 10^{19}$ cm$^{-3}$. This leads to a lower effective recombination rate constant, and resultant higher charge carrier densities and $V_{oc}$. It was noted the $V_{oc}$ was converging: Reducing the SRH recombination time constant by 2 orders of magnitude only reduced the initial $V_{oc}$ (739 mV) by 16 mV suggesting that it would not be possible to reach the experimentally observed initial value ($\approx 400$ mV). Furthermore, a negative TPV could not be obtained using a simulated light pulse (generating an additional $2.5 \times 10^{21}$ cm$^{-3}$s$^{-1}$ carriers). These results suggest that in the cells tested, $> 10^{17}$ cm$^{-3}$ ionic carriers are present. (**g**) and (**h**) Bottom cathode cell with $10^{18}$ cm$^{-3}$ mobile ionic defects equilibrated at short circuit (dashed lines) and been switched to open circuit (solid lines). In order to reach the experimentally observed $V_{oc}$, an increased SRH time constant of $\tau_{n,SRH} = \tau_{p,SRH} = 3 \times 10^{-16}$ s was required. However, all other results in this case remained similar to the case for $N_{ion} = 10^{19}$ cm$^{-3}$ due to shorter Debye lengths and effective screening of the built-in field.

**Note S2.** **Investigation of recombination schemes.**

Implementing SRH recombination ($\tau_{n,SRH} = 10^{-17}$ s) in a single contact (n-type) allowed the negative TPV transient to be reproduced (figure S6b blue trace) using the transient of the transient (TROTTR) protocol (figure 2a). The simulated $V_{oc}$ at time $t = 0$ was 200 mV higher than observed experimentally (figure S7a blue trace) and converging with respect to decreasing recombination time constant. This was due to the charge carrier concentrations in the contacts reaching their doping density equilibrium values. A number of factors could lead to this discrepancy in $V_{oc}$ such as lower-than-real rates of band-to-band recombination, a higher-than-real built-in voltage (which would alter the doping density in the contacts), and the uniform generation profile implemented in the simulation. The slow $V_{oc}$ transient also exhibited a characteristic hump (figure S7a blue trace) occasionally observed in top cathode $V_{oc}$ transients (figure S8b). Implementing high rates of band-to-band recombination ($k_{btb} = 10^{-2}$ cm$^3$s$^{-1}$) in all layers of the cell also enabled reproduction of the negative TPV at early times in the TROTTR (figure S7b purple trace). In this instance however, the slow $V_{oc}$ transient exhibited an initial negative deflection of around 120 mV, after which it plateaued approximately 100 mV above its initial value (figure S7a purple trace). Setting the bulk pervoskite band-to-band recombination rate to $10^{-1}$ cm$^3$s$^{-1}$ allowed the experimental $V_{oc}$ for the TROTTR experimental data to be obtained but neither the negative TPV nor slow $V_{oc}$ transient were reproduced (figure S7a and S7b red traces). Finally, switching from high rates of SRH to band-to-band recombination in the contacts produced very similar results to when SRH was implemented in the main findings (figure S7a and S7b yellow traces).

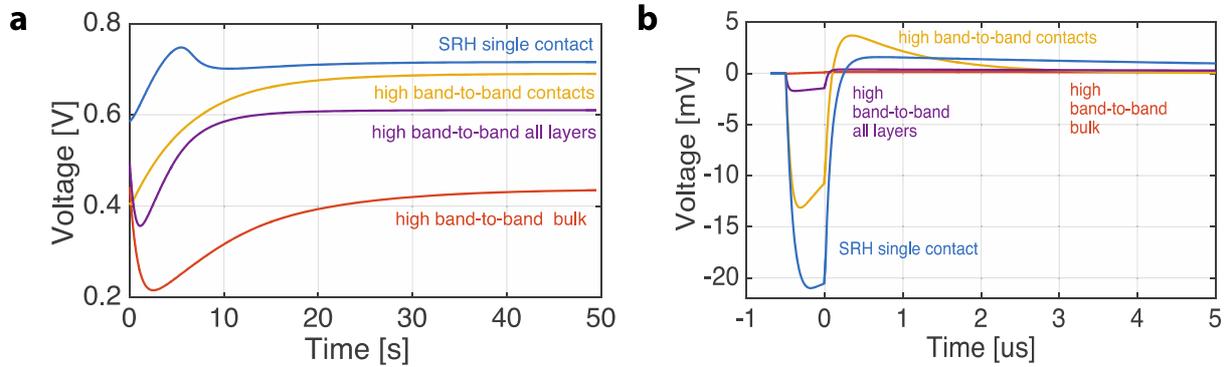

**Figure S8.** Voltage transient simulations under different recombination schemes. (**a**) Slow open circuit voltage transients after the cell had reached equilibrium at short circuit ($V_{\text{preset}} = 0$ V) and (**b**) initial photovoltage transients ($t = 0$ s) for four different recombination schemes trialled in the study. See note S2 for details.

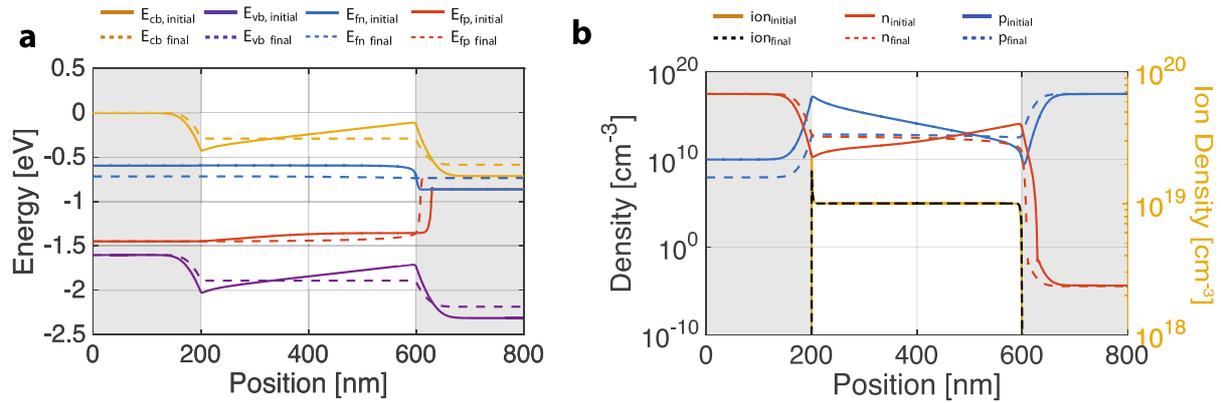

**Figure S9.** Single contact recombination simulation data. (**a**) Band diagram and (**b**) charge densities for simulated p-i-n structure solar cells with a high SRH recombination rate coefficient ($\tau_{n,\text{SRH}} = \tau_{p,\text{SRH}} = 2 \times 10^{-15}$ s) in the n-type contact only. Solid and dashed lines indicate initial and final states respectively.

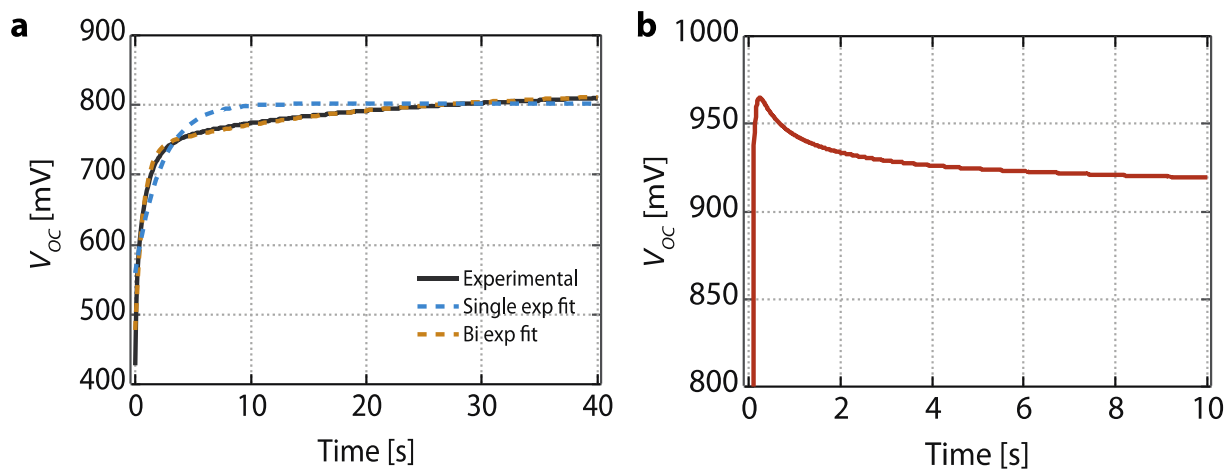

**Figure S10. Additional slow transient $V_{oc}$ experimental measurements.** Examples of biphasic increase in $V_{oc}$ with time for (**a**) a bottom cathode device after preconditioning at short circuit for 60 seconds before switching to open circuit and (**b**) a top cathode device. The characteristic hump in this case is suggestive of a higher rate of recombination in a single contact cf. figure S7a.

## Note S3. Simulation details

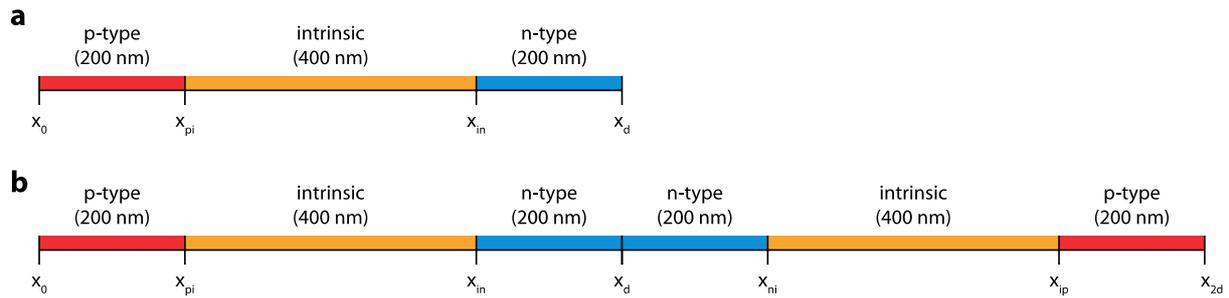

**Figure S11. Simulated device architectures.** Device architectures with position labels and layer thicknesses for (**a**) the fixed potential boundary condition p-i-n simulation used for JV and current transient measurements, and (**b**) the symmetric p-i-n-n-i-p cell used for simulating open circuit voltage transients.

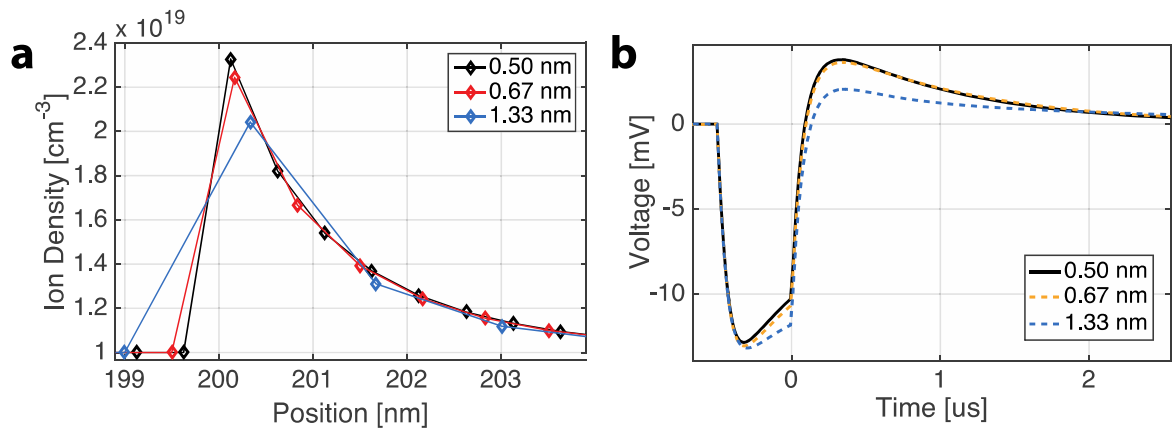

**Figure S10. Effect of mesh grid spacing on simulated solutions.** Solutions for a bottom cathode device undergoing a TROTR measurement are shown for grid spacings of 0.5, 0.67 and 1.33 nm. (**a**) Ion accumulation region at the p-type/perovskite interface at equilibrium and (**b**) transient photovoltage simulations after the cell has been equilibrated at short circuit then switched to open circuit with different mesh spacings. A spacing of 1.33 nm per points leads to large errors in regions of high charge density gradients, for example at the p-type/perovskite interface (figure S10a). The ion charge accumulation layer at this location has a characteristic length of a similar magnitude to the grid spacing (approximately 1.47 nm when the cell is at equilibrium) resulting in poor reproduction of the profile. These numerical inaccuracies are carried over to the transient photovoltage simulations (figure S10b). A higher resolution grid spacing of 0.5 nm per point leads to marginally more accurate solutions compared with the 0.67 nm spacing used for the simulations in the main text. It is clear however that the solutions are convergent with respect to reduced grid spacing. A spacing of 0.67 nm was chosen for all other simulations presented in this study as a compromise between numerical accuracy and calculation time.

**Physical model**

The device physics of the simulation is based on established semiconductor equations, detailed in Nelson[4] as described below.

**Table S1.    Key device simulation parameters.**

| Parameter name | Symbol | Value |
|---|---|---|
| Band gap[5] | $E_g$ | 1.6 eV |
| Built in voltage[6] | $V_{bi}$ | 1.3 V |
| Dielectric constant[7] | $\varepsilon_s$ | 20 |
| Mobile ionic defect density[8] | $N_{ion}$ | $10^{19}$ cm$^{-3}$ |
| Ion mobility | $\mu_a$ | $10^{-12}$ cm$^2$ V$^{-1}$ s$^{-1}$ |
| Electron mobility[9] | $\mu_e$ | 20 cm$^2$ V$^{-1}$ s$^{-1}$ |
| Hole mobility[9] | $\mu_h$ | 20 cm$^2$ V$^{-1}$ s$^{-1}$ |
| Nominal band-to-band recombination coefficient | $k_{btb}$ | $10^{-10}$ cm$^3$ s$^{-1}$ |
| p-type and n-type SRH time constants | $\tau_{n,SRH}, \tau_{p,SRH}$ | $2\times10^{-15}$ s* |
| SRH trap energy, cathode | $E_{t,\,ntype}$ | -0.2 eV (n-type)** |
| SRH trap energy, anode | $E_{t,\,ptype}$ | -1.4 eV (p-type) ** |
| Effective density of states | $N_0$ | $10^{20}$ cm$^{-3}$ |
| Generation rate | $G$ | $2.5\times10^{21}$ cm$^3$ s$^{-1}$ |

All values for the perovskite phase are roughly based on literature values, except the band-to-band recombination rate, which was calculated based on a homogeneous charge carrier density of $10^{16}$ cm$^{-3}$ and a nominal steady-state transient photovoltage decay time constant of 0.5 μs. The Shockley-Read-Hall (SRH) recombination rate coefficient and energies were chosen to reproduce the approximate open circuit voltage for the bottom cathode device. The generation rate was chosen to yield an absorbed photocurrent of ~ 16 mA cm$^{-2}$. The ionic mobility was adjusted from our previous literature estimate of about $4\times10^{-11}$ cm$^2$ V$^{-1}$ s$^{-1}$ (ref. [8]) to a lower value of $10^{-12}$ cm$^2$ V$^{-1}$ s$^{-1}$ to correctly approximate the evolution time of the transient.
* SRH coefficient used in bottom cathode device only.
** In the band gap, below conduction band (0 V)

**Table S2.** Table of Variables and symbols used for device simulation.

| Variable name | Symbol | Units |
|---|---|---|
| Electron charge carrier density | $n$ | $cm^{-3}$ |
| Hole charge carrier density | $p$ | $cm^{-3}$ |
| Mobile ionic defect charge carrier density | $a$ | $cm^{-3}$ |
| Intrinsic carrier density | $n_i$ | $cm^{-3}$ |
| n-type donor density | $N_D$ | $cm^{-3}$ |
| p-type acceptor density | $N_A$ | $cm^{-3}$ |
| Static ionic defect density | $N_{ion}$ | $cm^{-3}$ |
| n-type electron equilibrium density | $n_0$ | $cm^{-3}$ |
| p-type electron equilibrium density | $p_0$ | $cm^{-3}$ |
| Electrostatic potential | $V$ | V |
| Built in voltage | $V_{bi}$ | V |
| Applied voltage | $V_{app}$ | V |
| Current densities (electron, hole, ion defect) | $J_n, J_p, J_a$ | $A\,cm^{-2}$ |
| Generation rate | $G$ | $cm^3\,s^{-1}$ |
| Recombination rate | $U$ | $cm^3\,s^{-1}$ |
| Net free charge density | $\rho_f$ | $cm^{-3}$ |
| Position | $x$ | cm |
| Time | $t$ | s |

**Table S3**. Constants and values used for device simulation.

| Constant name | Symbol | Value (4 s.f.)* | Units |
|---|---|---|---|
| Boltzmann's constant | $k_B$ | $8.617 \times 10^{-5}$ | $eV\,K^{-1}$ |
| Temperature | $T$ | 300 | K |
| Vacuum permittivity | $\varepsilon_0$ | $5.524 \times 10^5$ | $q^2\,eV^{-1}\,cm^{-1}$ |
| Electron charge | $q$ | $1.619 \times 10^{-19}$ | C |

*Values in the simulation generally to higher accuracy.

**Simulation coupled equations**

The MATLAB pdepe 1-dimensional solver was used to solve the continuity equations and Poisson's equation for $n$, $p$, $a$ and $V$ as a function of time. The following relationships were used in the model.

**Total current density**

$$J = J_n + J_p + J_a + J_{disp} \tag{S10}$$

**Drift and diffusion current densities**

$$J_n = -q\mu_e \left(n\frac{dV}{dx} - k_B T \frac{dn}{dx}\right) \tag{S11}$$

$$J_p = q\mu_h \left(p\frac{dV}{dx} + k_B T \frac{dp}{dx}\right) \tag{S12}$$

$$J_a = q\mu_a \left(a\frac{dV}{dx} + k_B T \frac{da}{dx}\right) \quad x_{pi} < x < x_{in} \tag{S13}$$

$$J_a = 0 \quad 0 < x < x_{pi} \text{ \& } x_{in} < x < x_d \tag{S14}$$

**Displacement current density**

$$J_{disp} = \varepsilon_0 \varepsilon_s \frac{dE}{dt} \tag{S15}$$

**Continuity equations**

$$\frac{\partial n}{\partial t} = \frac{1}{q}\frac{\partial J_n}{\partial x} + G_n - U_n \quad 0 < x < x_d \tag{S16}$$

$$\frac{\partial p}{\partial t} = -\frac{1}{q}\frac{\partial J_p}{\partial x} + G_p - U_p \quad 0 < x < x_d \tag{S17}$$

$$\frac{\partial a}{\partial t} = -\frac{1}{q}\frac{\partial J_a}{\partial x} \quad x_{pi} < x < x_{in} \tag{S18}$$

$$\frac{\partial a}{\partial t} = 0 \quad 0 < x < x_{pi} \text{ \& } x_{in} < x < x_d \tag{S19}$$

**Poisson's equation**

$$\frac{\partial^2 V}{\partial x^2} = \frac{-q}{\varepsilon_0 \varepsilon_r}(n - p + N_A) \qquad 0 < x < x_{pi} \qquad (S20)$$

$$\frac{\partial^2 V}{\partial x^2} = \frac{-q}{\varepsilon_0 \varepsilon_r}(n - p - a + N_{ion}) \qquad x_{pi} < x < x_{in} \qquad (S21)$$

$$\frac{\partial^2 V}{\partial x^2} = \frac{-q}{\varepsilon_0 \varepsilon_r}(n - p - N_D) \qquad x_{in} < x < x_d \qquad (S22)$$

**Intrinsic and Doped Carrier Densities**

Intrinsic and doped equilibrium carrier densities were calculated using Boltzmann statistics. In order to generate the correct built-in potential at open circuit, doping densities in the *n* and *p*-type regions were calculated using the offset of the anode workfunction, $\phi_{ano}$ with the ionisation potential, $\phi_{IP}$, and the cathode workfunction, $\phi_{cat}$, with the electron affinity, $\phi_{EA}$. For simplicity, the contacts are assumed to be Ohmic.

$$n_i = N_0 \exp\left(\frac{-E_g}{2k_B T}\right) \qquad (S23)$$

$$N_D \approx n_0 = N_0 \exp\left(\frac{\phi_{cat} - \phi_{EA}}{k_B T}\right) \qquad (S24)$$

$$N_A \approx p_0 = N_0 \exp\left(\frac{\phi_{IP} - \phi_{ano}}{k_B T}\right) \qquad (S25)$$

**Open circuit voltage**

The open circuit voltage, $V_{OC}$ is given by the difference in electron and hole quasi Fermi levels at the n-type and p-type boundaries of the device respectively:

$$V_{oc} = V(x_d) - V(x_0) + \frac{k_B T}{q} \ln\left(\frac{n(x_d) p(x_0)}{n_i^2}\right) \qquad (S26)$$

**Recombination**

Band-to-band recombination, $U_{btb}$ in our model uses a simplified for of bimolecular recombination:

$$U_{btb} = k_{btb}(np - n_i^2) \tag{S27}$$

We used a reduced form of the Shockley-Read-Hall equation, where the capture cross section, mean thermal velocity, SRH coefficient and trap density are compressed into time constants, $\tau_{n,SRH}$ and $\tau_{p,SRH}$ for electrons and holes respectively.

$$U_{SRH} = \frac{np - n_i^2}{\tau_{n,SRH}(p+p_t) + \tau_{p,SRH}(n+n_t)}, \tag{S28}$$

where $n_t$ and $p_t$ are the electron and hole densities when the Fermi level is at the trap level.

**p-i-n Structure Boundary Conditions**

$$J_n(0) = 0 \tag{S29}$$

$$n(x_d) = n_0 \tag{S30}$$

$$p(0) = p_0 \tag{S31}$$

$$J_p(x_d) = 0 \tag{S32}$$

$$V(0) = 0 \tag{S33}$$

$$V(x_d) = V_{bi} - V_{app} \tag{S34}$$

**p-i-n-n-i-p Structure Boundary Conditions**

$$J_n(0) = 0 \tag{S35}$$

$$J_n(x_{2d}) = 0 \tag{S36}$$

$$J_p(0) = 0 \tag{S37}$$

$$J_p(x_{2d}) = 0 \tag{S38}$$

$$V(0) = 0 \tag{S39}$$

$$V(x_{2d}) = 0 \tag{S40}$$

**Supplementary References**